# Evaluation of Network Based IDS and Deployment of multi-sensor IDS

Snort, Suricata, Multi-Sensor IDS Deployment

Navya Iyengar
School of Computing & Digital Media, Robert Gordon University
Aberdeen, United Kingdom
n.iyengar@rgu.ac.uk

*Abstract*— **Cloud-based and network-based technology has witnessed an exponential rise in development. Adaptation of these latest technologies has opened flood gates for data breaches, an increase in sophistication of cyber threats, and, a multitude of new attack vectors. Numerous tools and solutions are currently available for detection of these threats. Network-based Intrusion Detection Systems (NIDS) are one of the most effective tools implemented to maintain confidentiality, integrity, and, availability of networks. Whilst there are several open-source tools in the offing, this paper evaluates two open-source NIDS – Snort and Suricata, along with strategic placement of multi-sensor IDS in a WAN environment, in combination with NIDS, for in time threat detection and protection of systems.**

*Keywords—Intrusion Detections Systems (IDS), Network Based IDS (NIDS), Snort, Suricata, multi-sensor IDS*

I. INTRODUCTION

Any unauthorized attempt of access or activity, in a network or computer, is classified as an intrusion. Combating these accesses, and, securing networks and system peripherals have led to the development of intrusion detection systems. The idea of implementing IDS is to detect novel attacks [1].

The new age e-business companies where revenues are dependent on web hosting or data hosting, any downtime in a network would lead to loss of revenue, customer base and fear of running out of business. IDS can be a device or a software application that is implemented to monitor and log, various malicious activity or policy violations, in networks or systems, thus making it an imperative tool. The complexity of IDS varies from being dedicated to a single system to large-complex network architecture [1].

IDS is classified based on its usage and implementation. Host-based IDS (HIDS) and Network-based IDS (NIDS).

   a. HIDS - It collects data on events in the form of audit trails maintained by the operating system of individual systems. The details available in audit trail make them a preferred source of data. They are scalable and cost-effective due to distributive property. Since they are host-based they have to be platform specific and do not support cross-platform functionality [2].

   b. NIDS – Data here is collected the network stream and follow the principle of "wiretapping". They are customisable basis organisation's network needs. They monitor only the network and are independent of the operating system. That also makes them financially viable. NIDS' is dynamic and can be altered as per requirement and usage [2].

   c. Architecture of NIDS in a network:

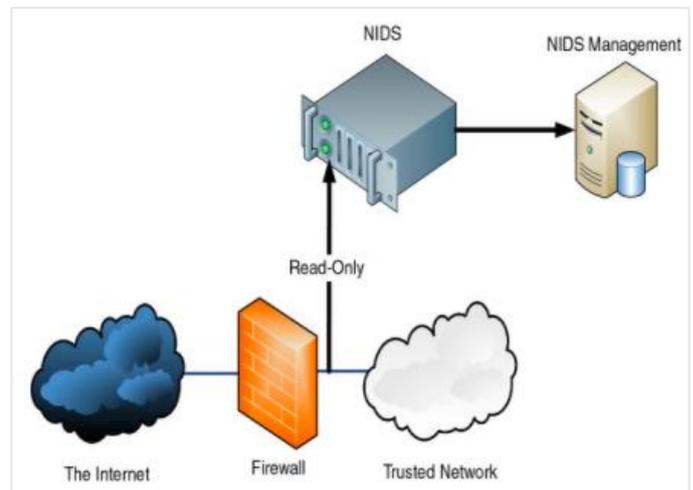

FIG 1 - NIDS ARCHITECTURE [11]

The objective is to compare and contrast 2 IDS technologies. The tools being evaluated are open source NIDS – Snort and Suricata.

II. OPEN SOURCE NETWORK-BASED IDS

*A. Snort*

Snort was created by Martin Roesch in 1998 [9]. One of the most widely deployed open source NIDS with a vast signature database, that performs real-time search on the network packet. It is compatible with almost all the operating systems (OS) like Mac OS, Linux, FreeBSD, Windows, Unix and Open BSD [3]. Snort can be configured and operated as both IDS and intrusion prevention system (IPS) [9].

The data coming from the network is captured and sent to the Packet Decoder module. Headers of these packets are monitored and later decrypted for further processing. Now in the pre-processor, it's put together with the TCP stream and HTTP URI is decrypted [9].

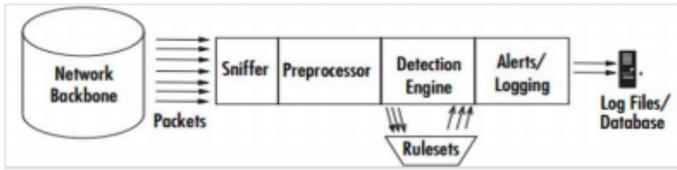

FIG 2 - SNORT NIDS ARCHITECTURE [12]

Detection engine verifies the packets against the rules applied and available in the tool. As a result, checks packets for a trace of any known attempt of intrusion. Regular packets are dropped while the doubtful ones are logged by the Logging and Alerting System. The Output Module accepts the logs and produces final output [3].

*B. Suricata*

Suricata is an open-source NIDS similar to Snort. It is an anomaly-based detection tool. The streams here are captured, decrypted, managed and finally analysed. It supports both pattern matching and scripts to detect attacks [4].

Once the data is captured in the capture module of the tool, Suricata makes it compatible for link type decoder. They are then decoded in Suricata supported data structure. The currently supported modules are LINKTYPE_LINUX_SLL, LINKTYPE_ETHERNET, LINKTYPE_PPP, LINKTYPE_RAW [9].

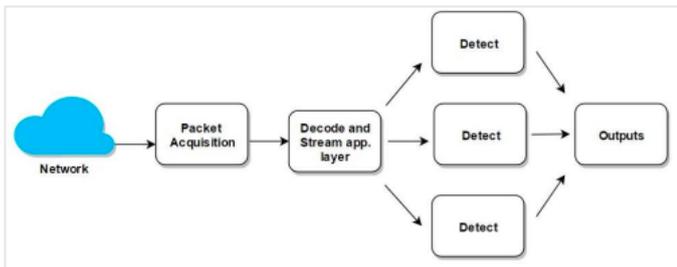

FIG 3 - SURICATA NIDS ARCHITECTURE [12]

Tasks like loading signatures, enabling detection plugins, forming detection groups for packet routing, subsequently running packets against the rules, are done by detection module [9].

*C. Previous Related Work*

Snort and Suricata NIDS have been compared by multiple researchers under various experimental conditions and network environment [5, 6, 7]. These research work included the performance of both the tools on various parameters, but not all paper had all the parameters. However, in this paper, the performance of the tools has been consolidated whilst pitting them against each other.

### III. COMPARISON AND ANALYSIS

Each tool is unique and the below table compares them basis various features.

| Parameters | Open Source NIDS | |
|---|---|---|
| | *Snort* | *Suricata* |
| Installation/Implementation | Simple | Complex |
| Operating System | OS agnostic | OS Agnostic |
| Threads | Single-threaded | Multi-threaded |
| Documentation and Support | Large community | Limited community |
| Cost Effective | Less expensive as works on system with lower configuration | Expensive as requires more system support |
| Network Speed | Slow since single threaded | Faster due to multi threads |
| Detction Method | Signature based | Anomaly based |
| Rule | Own VRT rule | Snort's VRT |

TABLE 1 – COMPARISON AND ANALYSIS TABLE

Whilst comparing different NIDS there are multiple parameters that it can be compared on. Table 1, lists a few of the attributes of both the NIDS.

*Rules* - Despite a similar set of rules, it's not entirely true. Snort uses subscriber rules, Suricata uses rules from emerging threats [5].

*RAM Usage* - Suricata uses more RAM power in comparison to Snort [6].

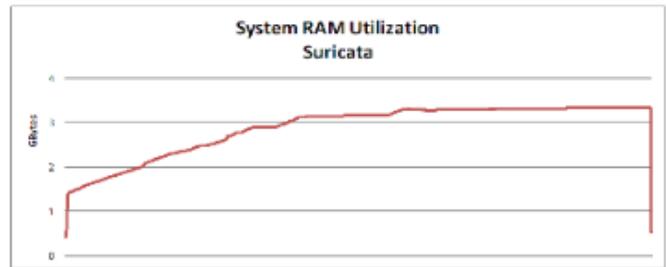

FIG 4 - SURICATA RAM USE [6]

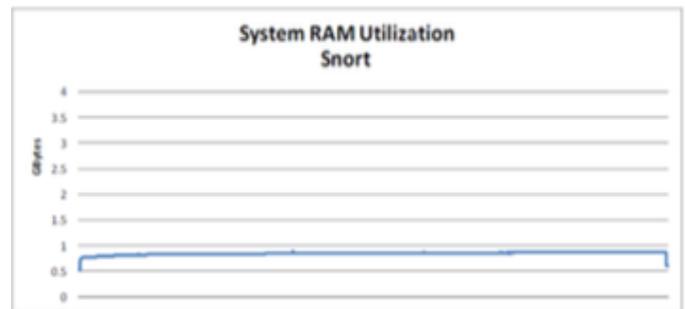

FIG 5 – SNORT RAM USE [6]

*CPU Utilisation* – The multi-threaded feature of Suricata uses more CPU in comparison to Snort for the same load of network monitoring [6].

*TCP Performance* – Snort is observed to have dropped packets in high-speed network which makes it difficult to use in Gbps speed networks, in comparison to Suricata which works best in high-speed networks [7].

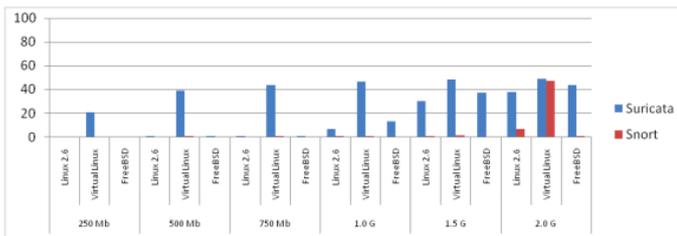

FIG 6 – TCP PERFORMANCE COMPARISON [7]

*UDP Performance* – Snort drops significantly lesser data packets in UDP (1047) [7].

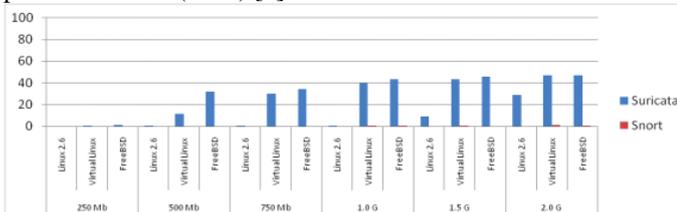

FIG 7 – UDP(1047) PERFORMANCE COMPARISON [7]

*Attack Detection Rate* – Since the packets are dropped by Suricata, the attack detection rate of Snort is extremely high at higher speeds [7].

*Ideal Platform* – Suricata as a tool best works on Linux OS in comparison to Snort, which works best on Linux in lower speeds and FreeBSD on higher speeds [7]. This doesn't limit their usage and can be deployed any OS.

## IV. MULTI-SENSOR IDS

Using multiple sensor IDS help in tuning each of them as per the network monitoring requirement, resulting in quicker identification and analysis of suspicious activities [8]. This would help achieve a robust solution of IDS. The optimal placement varies basis the requirement of the end result and the type of attack to be detected and is a network administrator's discretion.

Different IDS technologies (wireless, NBA – network behaviour analysis, network-based and host-based) have different architecture [10]. This is a study focused on the placement of multiple IDS sensors with a network-based IDS.

A NIDS sensor monitors and analyses network activity various network segments. The network interface card used for monitoring are placed in promiscuous mode, i.e., it will accept all inbound packets that they view, irrespective of their planned destinations. Most IDS deployments use multiple sensors, with deployments having a multitude of sensors [10].

Sensors can be deployed in 2 modes:
a. Inline Sensors – Actual network is made to pass through it if it has to be monitored, similar to a firewall. In order to achieve this, the NIDS sensor needs to be combined with the firewall. They are usually placed between the internal and external boundaries of the network. To prevent intrusion, it is best advised to have an inline mode of deployment [10].
b. Passive Sensors – A copy of the network traffic is monitored and there is no passage of the network involved. They are usually placed in a demilitarized zone (DMZ) subnet, a division between networks and other key points in a network [10].

### A. Deployment of Sensor

Sensors are to be placed at critical positions in a network which are considered sensitive and/or create a bottleneck.

a. WAN – Network in a WAN connection to a branch office is required to be watched [10].
b. External Network – The entire internal network connecting to the external network should be monitored for external threats and attacks [10].
c. Remote Access and Intranet – Internal network flow can be monitored [10].

### B. Placement Diagram

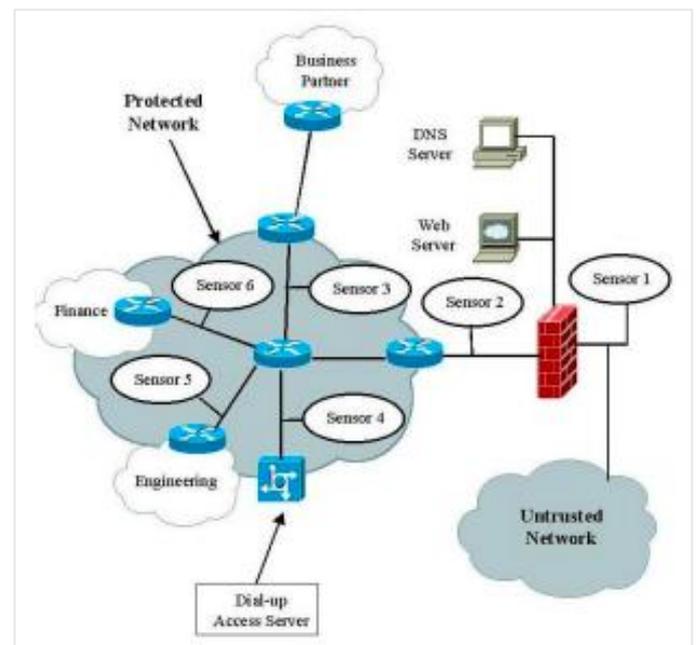

FIG 8 – SENSORS PLACEMENT IN NIDS [10]

Sensor 1 & 2 are placed strategically to monitor attacks from external sources. They guard the periphery of the network and act as an extended IDS sensor.

Sensor 3 & 4 are placed in order to monitor traffic flowing between the internal protected network and remote access given to workers working remotely.

Sensor 5 & 6 are to monitor network flow between the internal network assigned to various groups.

## V. CONCLUSION

This paper successfully compared and reviewed the features of two widely used open-source network-based intrusion detection system – Snort and Suricata. Both seem to be a very promising tool.

Despite Snort being used widely, it is worth noting that Suricata is a new age multi-threaded tool, which has a bigger scope of enhancements.

Another remarkable advantage Suricata has over Snort is, it doesn't require multiple instances to accommodate an increase in network traffic [6].

Subsequently, this paper also highlights how NIDS' can be used in combination with other types of IDS and firewall too to create a multi-sensor IDS. The deployment of such multi-sensor IDS is to be done basis the type of network or attacks to be watched. The suggestion to network administrator is basis the general network architecture and common network bottlenecks, this could vary with respect to network architecture of individual organization.